\documentclass[12pt]{article}
\usepackage{latexsym}
\usepackage[dvips]{graphics}
\usepackage{epsfig}
\usepackage{axodraw}
\usepackage{color}
\begin{document}
\begin{center}
{\bf \Large Conformal Invariance in the Leigh-Strassler\\[0.3cm]  deformed N=4
SYM Theory}\vspace{1cm}

{\bf \large L. V. Bork$^{2,4,\diamond}$, D. I. Kazakov$^{1,2,\dag}$, G. S. Vartanov$^{1,3,\ddag}$, \\[0.2cm]
and A. V. Zhiboedov$^{1,5,\star}$}\vspace{0.5cm}

{\it $^1$Bogoliubov Laboratory of Theoretical Physics, Joint
Institute for Nuclear Research, Dubna, Russia, \\
$^2$Institute for Theoretical and Experimental Physics, Moscow, Russia, \\
$^3$University Center, Joint Institute for Nuclear Research, Dubna, Russia,\\
$^4$Moscow Engineering Physics Institute, Moscow, Russia, \\
$^5$Moscow State University, Physics Department, Moscow,
Russia.}\vspace{1cm}

\abstract{We consider a full Leigh-Strassler deformation of the
${\cal N}=4$ SYM theory and look for conditions under which the
theory would be conformally invariant and finite. Applying the
algorithm of perturbative adjustments of the couplings we construct
a family of theories which are  conformal up to 3 loops in the
non-planar case and up to 4 loops in the planar one. We found
particular solutions in the planar case when the conformal condition
seems to be exhausted in the one loop order. Some of them happen to
be unitary equivalent to the real beta-deformed ${\cal N}=4$ SYM
theory, while others are genuine. We present the arguments that
these solutions might be valid in any loop order.}
\end{center}

\section{Introduction}

During the last decade much attention has been paid to the ${\cal
N}=4$ supersymmetric Yang-Mills theory (SYM) and its deformations
obtained by the orbifold~\cite{Bershadsky} or
orientifold~\cite{Armoni} projection, or by adding the marginal
deformations~\cite{Leigh} to the Lagrangian. Such deformations lead
to a theory with less supersymmetry but inheriting some attractive
features of the original ${\cal N}=4$ SYM theory, namely, the
conformal invariance, integrability~\cite{Integr1,Integr2} in the
planar limit, and, especially, its connection with\vfill
\hrule\vskip 1mm\noindent {$^{\diamond}$ e-mail: borkleonid@yandex.ru \\$^{\dag}$ e-mail: kazakovd@theor.jinr.ru \\
$^{\ddag}$ e-mail: vartanov@theor.jinr.ru} \\ $^{\star}$ e-mail:
zhiboedv@theor.jinr.ru \\ the dual string theory via the AdS/CFT
correspondence. This way it becomes possible to investigate
nonperturbative features of these theories.

Since the original version of the AdS/CFT
correspondence~\cite{Maldacena} there have appeared a lot of its
modifications~\cite{Zeytlin}. At the present time, it is not clear
how to build gravity dual to an arbitrary gauge theory or which
properties of the gauge theories are necessary for existence of this
correspondence. However it is obvious that conformal
invariance~\cite{fin} of the gauge theory plays a significant role
in this matter. As it was already mentioned, the Leigh-Strassler
deformation of the ${\cal N}=4$ SYM theory~\cite{Leigh} breaks the
initial supersymmetry to ${\cal N}=1$ supersymmetry and the
$SU(4)_R$ symmetry down to $U(1)_R$. One of such examples is the
so-called $\beta$-deformation of the original ${\cal N}=4$ SYM
theory. Its gravity dual was constructed by Lunin and
Maldacena~\cite{LM} and a significant role in this duality is played
by the $U(1) \times U(1)$ global symmetry of the $\beta$-deformed
theory which was associated with isometries of the deformed $AdS_5
\times \tilde{S}_5$ background. There are also attempts to construct
the gravity dual to the full Leigh-Strassler
deformation~\cite{Kulaxizi,AharKol,BerenCherkis}.

From the field theory side the investigation of the
$\beta$-deformation of the ${\cal N}=4$ SYM theory was dedicated
mainly to finding the conditions of conformal
invariance~\cite{Freedman,Zanon1,Zanon2,RSS} and
finiteness~\cite{Zanon2} of the theory, and to investigation of
Chiral Primary Operators(CPO)~\cite{RSS,Zanon4}. In the real $\beta$
case~\cite{Zanon1}, it was shown that the theory is exactly
conformal in the planar limit. For general $\beta$ the condition of
conformal invariance = finiteness in the planar limit was found up
to four loops in~\cite{KB}. In the nonplanar case, the conformal
condition was found up to three loops in~\cite{RSS} and recently the
first step towards the four-loop answer was made in~\cite{Ital}.

The case of the full Leigh-Strassler deformation was less
investigated from the quantum field theory side. The one-loop
conformal condition was obtained almost five years
ago~\cite{Razamat} while the three-loop anomalous dimension was
recently calculated in~\cite{Madhu} using the results of the
papers~\cite{Parkes,Jones}. Their result, however, seems not to
coincide with us and with the $\beta$-deformed case from~\cite{RSS}.
Also, some CPO were investigated in~\cite{Zanon4}. In this paper, we
look for the conformal invariance of the full Leigh-Strassler
deformation. Using the dimensional regularization(reduction) we
found conditions of conformal invariance up to four loops in the
planar limit and up to three loops in the non-planar one.

There are special cases when the conformal conditions are exhausted
in the one-loop order. In case of the $\beta$-deformed theory in the
planar limit, this corresponds to real values of $\beta$. We also
found such solutions for the full Leigh-Strassler deformation.
However, some of these solutions happen to be unitary equivalent to
the $\beta$-deformed case. This gives us a useful cross check of our
calculations. At the same time, also in the planar limit, there
exist non-trivial solutions which are not reduced to the
$\beta$-deformed ones. We present them below and conjecture that
they might be valid in any loop order.

This family of solutions does not possess any global symmetries,
except for $Z_3$, and has connections with the $\beta$-deformed
${\cal N}=4$ SYM at particular points. It would be very interesting
to understand their origin from the string theory side and build the
corresponding dual gravity background.

\section{The Leigh-Strassler Deformation of the ${\cal N}=4$ SYM Theory}

The so-called Leigh-Strassler deformation can be obtained by
modification of the superpotential in the original ${\cal N}=4$ SYM
theory written in terms of ${\cal N}=1$ superfields:
\begin{eqnarray}
  S&=& \int d^8z Tr\left(e^{-gV}\bar \Phi_i e^{gV} \Phi^i\right)
   +\left(\frac{1}{2g^2}\int d^6z Tr
  (W^\alpha W_\alpha) + \int d^6z\ {\cal W} + h.c.\right)
  \label{lag}
\end{eqnarray}
in such a way
\begin{eqnarray}
{\cal
W}_{N=4~SYM}&=&ig(Tr(\Phi_{1}\Phi_{2}\Phi_{3})-Tr(\Phi_{1}\Phi_{3}\Phi_{2}))\rightarrow
 \label{li-st_supp} \\
 {\cal W}_{LS~SYM
}&=&i[h_1Tr(\Phi_{1}\Phi_{2}\Phi_{3})-h_2Tr(\Phi_{1}\Phi_{3}\Phi_{2})
+\frac{h_{3}}{3}\sum_{i=1}^{3}Tr(\Phi_{i}^{3})], \nonumber
\end{eqnarray}
where $\Phi_i$ with $i=1,2,3$ are the three chiral superfields of
the original ${\cal N}=4$ SYM theory in the adjoint representation
of the gauge group $SU(N)$, and the couplings $h_1,h_2,h_3$ are in
general complex. The $\beta$-deformed case in the same notation
corresponds to
$$h_1=h q, \ h_2=h/q, \ q=e^{i\pi\beta} \ \mbox{and} \ h_3=0.$$

The Leigh-Strassler deformed superpotential breaks the $SU(4)_{R}$
symmetry of the original ${\cal N}=4$ theory down to $U(1)_{R}$. In
addition, it is invariant under cyclic permutations of
$(\Phi_1,~\Phi_2,~\Phi_3)$ and exchange: $\beta \leftrightarrow 1 -
\beta$ or in our notation $h_1 \leftrightarrow - h_2$.

In case of interest, as in any  ${\cal N}=1$ SYM theory formulated
in terms of ${\cal N}=1$ superfields, one  has two types of
divergent diagrams, those of the chiral field propagator and of the
gauge field one. The chiral vertices are finite due to the
non-renormalization theorems~\cite{non-ren} and for the gauge
vertices one can choose the background gauge~\cite{back} where their
divergent factors coincide with that of the gauge propagator. Thus,
the only divergent structures are the field propagators only.
Moreover, the gauge field propagator is not independent: its
divergent structure is related to the chiral field propagators. This
can be seen, for example, from the explicit form of the NSVZ gauge
beta function~\cite{NSVZ}  expressed in terms of the chiral field
anomalous dimensions $\gamma$ by
\begin{equation}\label{beta}
  \beta_g=g^2\frac{\sum T(R)-3C(G)-\sum T(R)\gamma(R)}{1-2gC(G)}, \ \ \ \ g\equiv
  g^2/16\pi^2.
\end{equation}
where $T(R)$ is the Dynkin index of a given representation $R$ and
$C(G)$ is the quadratic Casimir operator of the $SU(N)$ gauge group.
In the Leigh-Strassler deformed ${\cal N}=4$ SYM case one has the
same field content as in ${\cal N}=4$ SYM, so $\sum T(R)=3C(G)$ and
everything is defined by the chiral field anomalous dimension
$\gamma$. Since conformal invariance is understood as the vanishing
of the beta function, the Leigh-Strassler deformed theory is
(super)conformal invariant on the sub-manifold in the coupling
constant space which is defined by the following condition
\begin{eqnarray}\label{anom_dim_conf_cond}
\gamma(g,~\{h_{i}\})=0,
\end{eqnarray}
where $\{h_i\}=(h_1,~h_2,~h_3)$. One can solve this
condition~(\ref{anom_dim_conf_cond}) choosing the Yukawa couplings
in the form of perturbation series over $g$~\cite{finite}:
\begin{equation}\label{yuk}
  h_i=\alpha_{0i}g+ \alpha_{1i}g^3+ \alpha_{2i}g^5+...~,i=1...3.
\end{equation}
If the anomalous dimensions of the chiral fields vanish, so do the
gauge and Yukawa beta functions and the theory is conformally
invariant.

Conformal invariance also means that the theory is finite, i.e., all
UV divergencies cancel (or in some gauges the sum of divergencies)
and the renormalization factors $Z$ (or their products)  are equal
to 1  or finite. In the context of dimensional
regularization~\cite{scope} this can be achieved by adding to
expansion over $g$ (\ref{yuk}) a similar expansion over the
parameter of dimensional regularization $\varepsilon = 4-D$, i.e.,
one has the two-fold expansion instead of one-fold one~\cite{K1}
\begin{eqnarray}
  h_i&=&g\left(a_i+\alpha_{0i}^{(1)}\varepsilon+ \alpha_{0i}^{(2)}\varepsilon^2+...+
   \alpha_{0i}^{(n-2)}\varepsilon^{n-2}+\alpha_{0i}^{(n-1)}\varepsilon^{n-1}+ \alpha_{0i}^{(n)}\varepsilon^n+...\right) \nonumber\\
  &+&g^3\left(\alpha_{1i}^{(0)}+\alpha_{1i}^{(1)}\varepsilon+ \alpha_{1i}^{(2)}\varepsilon^2+...+
  \alpha_{1i}^{(n-2)}\varepsilon^{n-2}+\alpha_{1i}^{(n-1)}\varepsilon^{n-1}+...\right) \nonumber\\
  &+&g^5\left(\alpha_{2i}^{(0)}+\alpha_{2i}^{(1)}\varepsilon+ \alpha_{2i}^{(2)}\varepsilon^2+...+
  \alpha_{2i}^{(n-2)}\varepsilon^{n-2}+...\right) \nonumber\\
&+& ................  \nonumber \\
&+&g^{2n-1}\left(\alpha_{n-2i}^{(0)}+\alpha_{n-2i}^{(1)}\varepsilon+
.....\right) \nonumber\\
&+&g^{2n+1}\left(\alpha_{n-1i}^{(0)}+...\right).\label{hi}
\end{eqnarray}
In a given order of PT equal to $n$ one needs all terms of the
double expansion with a total power of $g^2\cdot \varepsilon$ equal
$n$. The existing freedom of choice of the coefficients
$\alpha_{ki}^{(m)}$ is sufficient to get {\it simultaneously} the
vanishing of the anomalous dimensions (read {\it conformal
invariance}) and the pole terms in $Z$ factors (read {\it
finiteness}). The coefficients from $\alpha_{ni}^{(0)}$ to
$\alpha_{0i}^{(n)}$ calculated in the $n$-th order of PT are
related. One cannot put either of them to zero in an arbitrary way.
For a more complete discussion and some examples of how these
procedure works see our previous paper~\cite{KB}.

Our goal now is to calculate several terms of the double
expansion~(\ref{hi}) and to look for particular solutions when
expansion breaks down at the first terms. In the case of a
$\beta$-deformed SYM theory such a solution was found
in~\cite{Zanon2} and corresponds to the real deformations, i.e., to
$|q|=1$.

\begin{figure}[t]
\fcolorbox{white}{white}{
  \begin{picture}(450,127) (2,-36)
    \SetWidth{0.8}
    \Line(153,52)(278,52)
    \Line(2,52)(127,52)
    \Line(306,52)(431,52)
    \CArc(212,46.6)(35.41,8.77,171.23)
    \PhotonArc(63.12,46.73)(35.27,8.6,174.67){-7.5}{5.5}
\SetColor{Red}\Vertex(95,51){4}\SetColor{Black}
    \Vertex(247,52){4}
    \Vertex(177,51){4}
    \CArc(370,46.6)(35.41,8.77,171.23)
\SetColor{Green} \Vertex(405,51){4} \Vertex(336,51){4}
\SetColor{Black}
    \CArc(212,-6)(27.31,114,474)
    \Line(239,-8)(272,-8)
    \Line(151,-8)(184,-8)
\SetColor{Red}\Vertex(32,51){4}\SetColor{Black}
  \Text(130,-8)[]{ $\mathbf \Longrightarrow$}
  \Text(159,60)[]{$\bar{\Phi}_{i}$}
  \Text(7,60)[]{$\bar{\Phi}_{i}$}
  \Text(125,60)[]{$\Phi_{i}$}
  \Text(274,60)[]{$\Phi_{i}$}
  \Text(312,60)[]{$\bar{\Phi}_{i}$}
  \Text(428,60)[]{$\Phi_{i}$}
  \Text(40,43)[]{$\bar{D}^{2}$}
  \Text(87,43)[]{$D^{2}$}
  \Text(187,43)[]{$D^{2}$}
  \Text(238,43)[]{$\bar{D}^{2}$}
  \Text(349,43)[]{$D^{2}$}
  \Text(396,43)[]{$\bar{D}^{2}$}
  \end{picture}}
\caption{Supergraphs contributing to the chiral propagator at 1
loop and their scalar counterpart.  }\label{graph1}
\end{figure}
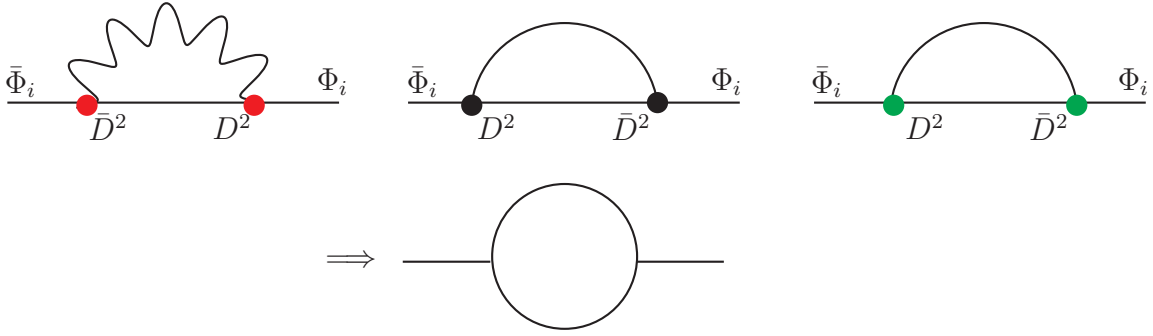

In dimensional regularization (reduction) and $\overline{MS}$
renormalization scheme the anomalous dimension of a chiral
superfield has the following form in the n-th order of PT:
\begin{eqnarray}\label{gamma}
\gamma(g,~\{h_{i}\})=\sum_{k=1}^{n}k~c_{1k}(g,~\{h_{i}\}),
\end{eqnarray}
where $c_{1k}$ are the coefficients at the lowest order pole in
$Z_{2}^{-1}$. In the 1-loop order one has for the chiral field
renormalization constant
\begin{equation}\label{1loopZ}
Z_{2}^{-1}=1-\frac{N}{(4\pi)^2}\left(f(\{h_{i}\},N)-2g^2\right)\frac{1}{\varepsilon}.
\end{equation}
Contributions to $Z_{2}^{-1}$ are presented in Fig.\ref{graph1}
where red, black, and green dots correspond to chiral-gauge
$\bar{\Phi}V\Phi$, chiral $h_1,~h_2$ and chiral $h_3$ vertices.
After performing D-algebra all diagrams in Fig.\ref{graph1} reduce
to the same scalar logarithmically divergent integral with different
colour factors ( hereafter we used SusyMath ver. 1.1~\cite{SusyMath}
and FeynCalc 5.1~\cite{FeynCalc} \emph{Mathematica} packages to
verify our calculations ). From~(\ref{1loopZ}) one can see that
\begin{equation}\label{1loopc}
c_{11}=-\frac{N}{(4\pi)^2}\left(f(\{h_{i}\},N)-2g^2\right),
\end{equation}
where
\begin{equation}\label{1loopg}
f(\{h_{i}\},N)=\sum_{i,k=1}^{3}f_{ik}h_{i}\bar{h}_{k}=(1-\frac{2}{N^2})(|h_1|^2
+ |h_2|^2) + \frac{2}{N^2} (h_{1}\bar{h}_{2} + h_{2}\bar{h}_{1}) +
(1-\frac{4}{N^2})|h_3|^2,
\end{equation}
so the  nonzero coefficients $f_{ik}$ are
\begin{equation}\label{1loopf}
f_{11}=f_{22}=(1-\frac{2}{N^2}),~f_{33}=(1-\frac{4}{N^2}),~f_{21}=f_{12}=\frac{2}{N^2},
\end{equation}
where $N$ is the number of colors of the gauge group $SU(N)$.

Thus, the one-loop conformal condition takes the form
\begin{equation}\label{1conf}
f(\{h_{i}\},N)-2g^2=0.
\end{equation}
To fulfil it, the coefficients in the expansion~(\ref{hi})
$\{a_i\}=(a_1,~a_2,~a_3)$ must then satisfy the following
requirement:
\begin{equation}\label{1loopa}
\sum_{i,k=1}^{3}f_{ik}a_{i}\bar{a}_{k}=2.
\end{equation}
To find other terms of expansion~(\ref{hi}), one has to calculate
the pole coefficients $c_{ik}$ of (\ref{gamma}) at higher orders of
PT. For simplicity, we consider everywhere only the difference
between the Leigh-Strassler deformed and the undeformed ${\cal N}=4$
SYM theory since calculating the difference we skip the calculation
of many diagrams with gauge lines inside the diagrams~\cite{RSS}.
The resulting expressions have some common structure in all orders
of PT which simplifies the analysis:

Up to three loops in the planar case (or up to two loops in the
non-planar case) the coefficients $c_{ik}$ have the following form:
\begin{equation}\label{coef}
c_{nk}=(f(\{h_{i}\},N)-2g^2)P_{nk}(h_{i},g^2,N),~n=1,..,3,\
k=1,...,n~,
\end{equation}
where $P_{nk}(h_{i},g^2,N)$ is a homogenous polynomial of the
form:
\begin{equation}
P_{nk}(\{h_{i}\},g^2,N)=\sum_{L=0}^{n-1}\sum^{3}_{i,k=1}
(P_{nk})_{ikL}(h_{i}\bar{h}_{k})^{L}(g^2)^{(n-1)-L},~k=1,...,n,
\end{equation}
where $(P_{nk})_{ikL}$ are some real numbers. One can see that the
one-loop conformal condition~(\ref{1conf}) is exact up to 3 loops
in the planar case and up to two loops in the non-planar case. In
higher orders new contributions appear and eq.(\ref{coef}) is
modified.

\subsection{Three-Loop (Non-Planar Limit) conformal condition}

\begin{figure}[t]
\fcolorbox{white}{white}{
  \begin{picture}(430,142) (-1,0)
    \SetWidth{0.8}
    \Line(-1,58)(50,58)
    \CArc(104,58)(53.74,135,495)
    \Line(158,58)(200,58)
    \Line(242,58)(283,58)
    \CArc(337,58)(53.74,135,495)
\Line(391,58)(422,58)
    \Line(283,58)(391,58)
    \Line(137,101)(113,67)
    \Line(336,58)(336,112)
    \Line(74,13)(98,47)
    \CArc(109.18,54.87)(13.67,73.8,215.13)
    \Line(73,102)(135,13)
    \Text(220,58)[]{ $\mathbf \Longrightarrow$}
   \Text(3,65)[]{$\bar{\Phi}$}
   \Text(198,64)[]{$\Phi$}
   \Text(43,65)[]{$D^{2}$}
   \Text(64,105)[]{$\bar{D}^{2}$}
   \Text(80,113)[]{$\bar{D}^{2}$}
   \Text(80,1)[]{$\bar{D}^{2}$}
   \Text(65,9)[]{$\bar{D}^{2}$}
   \Text(150,103)[]{$D^{2}$}
   \Text(138,111)[]{$D^{2}$}
   \Text(132,3)[]{$D^{2}$}
   \Text(147,14)[]{$D^{2}$}
   \Text(167,65)[]{$\bar{D}^{2}$}
  \end{picture}
} \caption{The topology of the relevant divergent non-planar
supergraphs and their scalar counterpart at 3 loops}\label{graph2}
\end{figure}
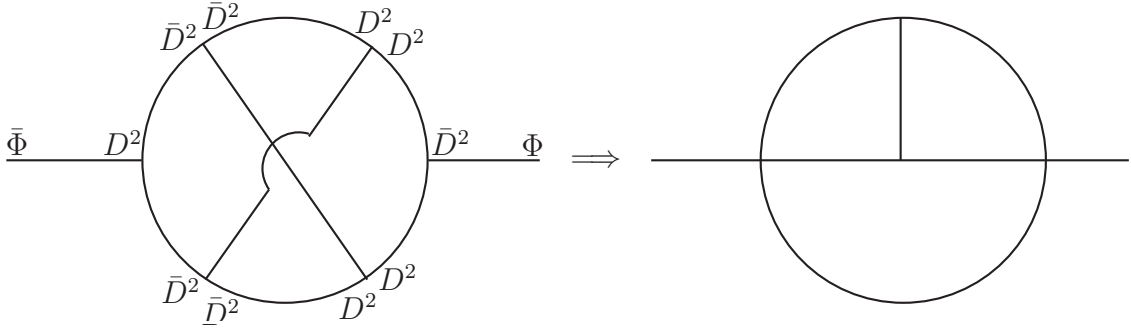

Starting from three loops in the nonplanar case one has the new
contribution coming from the set of supergraphs with the "cross"
topology (see Fig.2). Equation(\ref{coef}) then takes the form
\begin{equation}
c_{nk}=(f(\{h_{i}\},N)-2g^2)P_{nk}(\{h_{i}\},g^2,N)+G_{nk}(\{h_{i}\},N),~n
\geq 3,\ k=1,...,n~,
\end{equation}
where
\begin{equation}
G_{nk}(\{h_{i}\},N)=\sum^{3}_{i,p=1}(G_{nk})_{ip}(h_{i}\bar{h}_{p})^{n},
\end{equation}
is a homogeneous polynomial, and
\begin{equation}
G_{nk}(\{a_{i}g\},N) \neq 0,
\end{equation}
i.e., $G_{nk}$ do not vanish when applying the one loop conformal
condition~(\ref{1conf}), and to achieve conformal invariance one has
to take more terms of the double expansion (\ref{hi}). At this order
of PT, to get simultaneously conformal and finite theory, one needs
the following terms of expansion~(\ref{hi}):
\begin{eqnarray}\label{hi_1}
h_i &=& g \left( a_1 + \alpha_{0i}^{(2)} \varepsilon^2 +
g^2\alpha_{2i}^{(1)} \varepsilon^1 + g^4 \alpha_{4i}^{(0)}
\right),~i=1,2,3.
\end{eqnarray}
The only nonvanishing  contribution   at this order of PT is
$G_{31}$.  The explicit form of $G_{31}$ comes from the set of three
loop nonplanar supergraphs with "cross" topology~(Fig.\ref{graph2}).
The D-algebra for every supergraph in this set is identical and
leads to the same bosonic integral.  It is easy to see that every
supergraph with "cross" topology has no divergent subgraphs and
every such supergraph contributes only to the simple pole
coefficient in the singular part of the bare chiral propagator
$\langle\Phi\bar{\Phi}\rangle_{B}$. So $G_{31}=-D_{31}$, where
$D_{31}$ is the pole coefficient in the singular part of the
$\langle\Phi\bar{\Phi}\rangle_{B}$, and looks like
\begin{eqnarray}
G_{31}(\{h_{i}\},N)&=&
-\frac{1}{128}\frac{6\zeta(3)}{(4\pi)^6}\frac{N^2-4}{N^3}
\times \\
&&\hspace*{-3cm}  \left\{ |h_1-h_2|^2 \left(
N^2|h_{1}^2+h_{2}^2+h_{1}h_{2}|^2 - 9N^2|h_{1}|^2|h_{2}|^2 +
5|h_{1}-h_{2}|^4\right)\right.
\nonumber\\
&&\hspace*{-3cm} -18 |h_3|^2 \left( (N^2-5)|h_{1}^{2}+h_{2}^{2}|^2
- (N^2-10) \left(\bar{h}_{1}\bar{h}_{2}(h_{2}^{2}+h_{1}^{2}) +
c.c.\right) - 20|h_{1}|^{2}|h_{2}|^{2}\right)
\nonumber\\
&&\hspace*{-3cm} +\left( \bar{h}_{3}^{3}(h_{1}-h_{2})(
(N^2+20)(h_{1}^{2}+h_{2}^{2}) + 10(N^2-4)h_{1}h_{2}) + c.c.\right)
\nonumber\\
&&\hspace*{-3cm}\left. - 8 (N^2-10)(|h_3|^2)^3 \right\}\label{g31}
.
\end{eqnarray}

Now we follow the standard procedure~\cite{KB}: from the
requirement of vanishing of the anomalous dimension one has up to
3 loops:
\begin{eqnarray}\label{3loop_anom_dim}
\gamma=c_{11}+2c_{21}+3c_{31}=0
\end{eqnarray}
and substituting (\ref{hi_1}) in (\ref{3loop_anom_dim}) one  has
\begin{eqnarray}
1\ loop: & &\ \sum_{i,k=1}^{3}f_{ik}a_{i}\bar{a}_{k}=2, \label{vanNP} \\
3\ loops: & &
d_{1}\sum_{i,k=1}^{3}f_{ik}(a_{i}\bar{\alpha}^{(0)}_{4k}+\alpha^{(0)}_{4i}\bar{a_{k}})
 = - 3d_2G_{31}^{\Sigma}, \nonumber
\end{eqnarray}
where hereafter we define
\begin{eqnarray}
G_{31}(\{a_{i}g\},N)=d_2G_{31}^{\Sigma}g^6.
\end{eqnarray}
and $d_1=\frac{N}{(4\pi)^2},\ \
d_2=-\frac{N^3}{128}\frac{6\zeta(3)}{(4\pi)^6}$.
 The explicit form of
$G_{31}^{\Sigma}$ is:
\begin{eqnarray}\label{G31non}
G_{31}^{\Sigma}&=& \frac{N^2-4}{N^6}\{|a_1-a_2|^2 \left(
N^2|a_{1}^2+a_{2}^2+a_{1}a_{2}|^2 - 9N^2|a_{1}|^2|a_{2}|^2 +
5|a_{1}-a_{2}|^4\right)
\nonumber\\
&&\hspace*{-2cm} -18 |a_3|^2 \left( (N^2-5)|a_{1}^{2}+a_{2}^{2}|^2
- (N^2-10) \left(\bar{a}_{1}\bar{a}_{2}(a_{2}^{2}+a_{1}^{2}) +
c.c.\right) - 20|a_{1}|^{2}|a_{2}|^{2}\right)
\nonumber\\
&&\hspace*{-2cm} +\left( \bar{a}_{3}^{3}(a_{1}-a_{2})(
(N^2+20)(a_{1}^{2}+a_{2}^{2}) + 10(N^2-4)a_{1}a_{2}) + c.c.\right)
\nonumber\\
&&\hspace*{-2cm} -8 (N^2-10)(|a_3|^2)^3\}.
\end{eqnarray}

To get $\alpha_{0i}^{(2)}$, according to~\cite{KB}, one has to
consider $\langle\Phi\bar{\Phi}\rangle_{B}$. From the requirement of
vanishing of all poles in $\langle\Phi\bar{\Phi}\rangle_{B}$ one has
\begin{equation}\label{two}
6d_{1}^{3}\sum_{i,k=1}^{3}f_{ik}(a_{i}\bar{\alpha}_{0k}^{(2)}
+\alpha_{0i}^{(2)}\bar{a}_{k})-d_2G_{31}^{\Sigma}=0.
\end{equation}
We used the RG equations to restore the necessary higher pole
coefficients. To reach the total finiteness, one can use the
remaining coefficients. From the requirement that $Z_{2}^{-1}=1$ in
3 loops one gets, as in~\cite{KB},
\begin{eqnarray}&&
3d_{1}^{2}\sum_{i,k=1}^{3}f_{ik}(a_{i}\bar{\alpha}_{2k}^{(1)}
+\alpha_{2i}^{(1)}\bar{a}_{k}) + d_{1}\sum_{i,k=1}^{3}f_{ik}(a_{i}
\bar{\alpha}^{(0)}_{4k}
+\alpha^{(0)}_{4i}\bar{a_{k}}) \nonumber \\
&&+6d_{1}^{3}\sum_{i,k=1}^{3}f_{ik}(a_{i}\bar{\alpha}_{0k}^{(2)}
+\alpha_{0i}^{(2)}\bar{a}_{k})g^6 + d_2G_{31}^{\Sigma}=0,
\end{eqnarray}
or using (\ref{vanNP},\ref{two}):
\begin{equation}
3d_{1}^{2}\sum_{i,k=1}^{3}f_{ik}(a_{i}\bar{\alpha}_{2k}^{(1)}
+\alpha_{2i}^{(1)}\bar{a}_{k})-d_2G_{31}^{\Sigma}=0.
\end{equation}
Putting all together we obtain that up to 3 loops $\{h_{i}\}$ must
satisfy the following condition:
\begin{eqnarray}\label{43lconf}
\sum_{i,k=1}^{3}f_{ik}h_{i}\bar{h}_{k}&=&(1-\frac{2}{N^2})(|h_{1}|^2+|h_{2}|^2)
+\frac{2}{N^2}(h_{1}\bar{h}_{2}+h_{2}\bar{h}_{1})+(1-\frac{4}{N^2})|h_3|^2
\\
&&\hspace*{-3cm} =g^{2}\left\{2-\frac{\zeta_3}{128}G_{31}^{\Sigma}
\varepsilon^{2} - \frac{2\zeta_3}{128}G_{31}^{\Sigma}
\left(\frac{g^2N}{16\pi^2}\right)\varepsilon +
\frac{18\zeta_3}{128}G_{31}^{\Sigma}
\left(\frac{g^2N}{16\pi^2}\right)^2\right\}\nonumber
\end{eqnarray}
For the bare couplings one has:
\begin{eqnarray}
\sum_{i,k=1}^{3}f_{ik}(h_{i}
\bar{h}_{k})|_{B}&=&g^{2}_{B}\left\{2-\frac{\zeta_3}{128}G_{31}^{\Sigma}\varepsilon^{2}+...\right\}
\end{eqnarray}
For any values of the coefficients in~(\ref{hi_1}) which satisfy
(\ref{43lconf})  the theory is conformally invariant and finite up
to three loops. In the planar limit we see from(\ref{G31non}) that
the coefficient $G_{31}^{\Sigma}$ vanishes, which leads us to the
one-loop conformal condition.

\subsection{Four-Loop (Planar Limit) conformal condition}

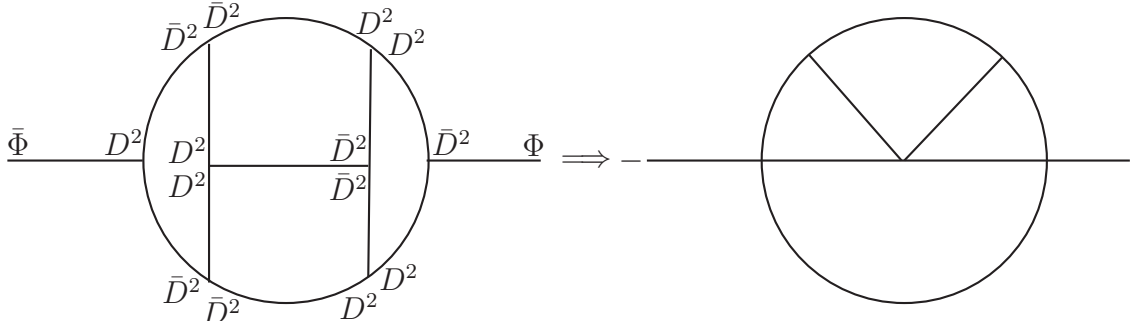
\begin{figure}[t]
\fcolorbox{white}{white}{
  \begin{picture}(430,117) (-1,-25)
    \SetWidth{0.8}
    \Line(-1,33)(50,33)
    \CArc(104,33)(53.74,135,495)
    \Line(157,33)(200,33)
    \Line(240,33)(283,33)
    \CArc(337,33)(53.74,135,495)
\Line(391,33)(422,33)
    \Line(75,-13)(75,77)
    \Line(75,31)(135,31)
    \Line(283,33)(391,33)
    \Line(336,33)(301,73)
    \Line(337,33)(374,72)
    \Line(135,-11)(136,75)
   \Text(216,33)[]{ $\mathbf \Longrightarrow$}
   \Text(233,33)[]{ $\mathbf -$}
   \Text(3,40)[]{$\bar{\Phi}$}
   \Text(198,39)[]{$\Phi$}
   \Text(43,40)[]{$D^{2}$}
   \Text(64,80)[]{$\bar{D}^{2}$}
   \Text(80,88)[]{$\bar{D}^{2}$}
   \Text(67,37)[]{$D^{2}$}
   \Text(67,24)[]{$D^{2}$}
   \Text(80,-24)[]{$\bar{D}^{2}$}
   \Text(65,-16)[]{$\bar{D}^{2}$}
   \Text(150,78)[]{$D^{2}$}
   \Text(138,86)[]{$D^{2}$}
   \Text(128,38)[]{$\bar{D}^{2}$}
   \Text(128,23)[]{$\bar{D}^{2}$}
   \Text(132,-22)[]{$D^{2}$}
   \Text(147,-11)[]{$D^{2}$}
   \Text(167,40)[]{$\bar{D}^{2}$}
  \end{picture}
}
\caption{The topology of the relevant divergent planar supergraphs
and their scalar counterpart at 4 loops}\label{graph3}
\end{figure}

The situation is simplified in the planar ( large $N$ of the $SU(N)$
gauge group ) limit. In this case, in the one-loop conformal
condition~(\ref{1loopg}) only the diagonal terms $f_{ik},~i=k$
survive
\begin{equation}
f(\{h_{i}\},N)=\sum_{i,k=1}^{3}f_{ik}h_{i}\bar{h}_{k}=|h_1|^2 +
|h_2|^2+|h_3|^2,
\end{equation}
so from~(\ref{1conf}) one has
\begin{equation}\label{1loopg}
|h_1|^2 + |h_2|^2+|h_3|^2-2g^{2}=0.
\end{equation}
At three loops all $G_{ik}=0$ (note that the set of supergraphs with
"cross" topology does not survive in the planar limit). At four
loops the only  nonvanishing contribution to $G_{41}$ comes from the
set of planar supergraphs with the new "ladder" topology (see
Fig.\ref{graph3}). The D-algebra for every supergraph in this set is
identical and leads to the same bosonic integral.  It is easy to see
that every chiral supergraph with the "ladder" topology has no
divergent subgraphs. The contribution of this set of chiral
supergraphs to the chiral propagator renormalization constant in the
planar limit is:
\begin{eqnarray}
c_{41}(\{h_{i}\},g^2,N)=  \label{g41}\\
&&\hspace*{-3cm}
=\frac{5}{2}\zeta(5)\frac{N^{4}}{(4\pi)^{8}}\{(|h_1|^2+|h_2|^2+|h_3|^2)^{4}-(2g^{2})^{4}
+(|h_1|^{2}-|h_2|^2)^4 + (|h_3|^2)^4 \nonumber \\ &&\hspace*{-3cm}
+6(|h_3|^2)^2 (|h_1|^2+|h_2|^2)^2 + 24|h_3|^2 |h_1|^2 |h_2|^2
(|h_1|^2+|h_2|^2) + \nonumber
\\ &&\hspace*{-3cm} +8 h_3^3 (|h_2|^2 \bar h_1^3 - |h_1|^2 \bar h_2^3) + 8 \bar h_3^3 (|h_2|^2 h_1^3 - |h_1|^2 h_2^3) \nonumber \\
&&\hspace*{-3cm} -8 |h_3|^2 (h_2^3 \bar h_1^3 + h_1^3 \bar h_2^3)
- 4 |h_3|^2 (|h_1|^2+|h_2|^2)^3 - 4 (|h_3|^2 )^3
(|h_1|^2+|h_2|^2)\} .\nonumber
\end{eqnarray}
Hereafter the chiral-gauge $\bar{ \Phi } V \Phi$ contributions
proportional to $|h_1|^2+|h_2|^2+|h_3|^2-2g^{2}$ are omitted. Note
that in this case $G_{41}=c_{41}$ and does not vanish at the
one-loop conformal condition.

With account of nonvanishing contribution to $G_{41}$ one needs the
following terms of expansion~(\ref{hi}):
\begin{eqnarray}\label{h1}
h_i &=& g \left( a_i + \alpha_{0i}^{(3)} \varepsilon^3 +
g^2\alpha_{2i}^{(2)} \varepsilon^2 + g^4 \alpha_{4i}^{(1)}
\varepsilon + g^6 \alpha_{6i}^{(0)} \right),~i=1,2,3.
\end{eqnarray}
From the requirement of vanishing of the anomalous dimension
$\gamma=c_{11}+2c_{21}+3c_{31}+4c_{41}=0$, one finds
\begin{eqnarray}
1\ loop: & &\ \sum_{i,k=1}^{3}f_{ik}a_{i}\bar{a}_{k}= \ 2, \label{van} \\
4\ loops: & & d_1[( \overline{a}_1 \alpha_{31}^{(0)} + a_1
\overline{\alpha}_{31}^{(0)}) + ( \overline{a}_2 \alpha_{32}^{(0)}
+ a_2 \overline{\alpha}_{32}^{(0)}) + ( \overline{a}_3
\alpha_{33}^{(0)} + a_3 \overline{\alpha}_{33}^{(0)}) ]=
-4d_2G_{41}^\Sigma, \nonumber
\end{eqnarray}
where now $ d_{1}=\frac{N}{(4\pi)^{2}},~
d_2=\frac{5}{2}\frac{\zeta(5)N^{4}}{(4\pi)^{8}}$. The explicit
form of $G_{41}^\Sigma$ is:
\begin{eqnarray}\label{G}
G_{41}^{\Sigma} &=& \large{\{} (a_3\overline{a}_3)^4 +
(a_1\overline{a}_1-a_2\overline{a}_2)^4 +
6(a_3\overline{a}_3)^2(a_1\overline{a}_1+a_2\overline{a}_2)^2
\nonumber \\ &+&
24a_1\overline{a}_1a_2\overline{a}_2a_3\overline{a}_3(a_1\overline{a}_1+a_2\overline{a}_2)
+ 8a^3_3(a_2\overline{a}_2\overline{a}_1^3-a_1\overline{a}_1\overline{a}^3_2) \nonumber\\
& +&
8\overline{a}^3_3(a_2\overline{a}_2a^3_1-a_1\overline{a}_1a^3_2) -
8a_3\overline{a}_3(\overline{a}_1^3a^3_2+\overline{a}_2^3a^3_1) \nonumber \\
&-& 4 ca_3\overline{a}_3 (a_1\overline{a}_1+a_2\overline{a}_2)^3 -
4(a_3\overline{a}_3)^3(a_1\overline{a}_1+a_2\overline{a}_2)
\large\}.
\end{eqnarray}

To get $\alpha_{0i}^{(3)}$, according to~\cite{KB}, one has to
consider the bare propagator. Since the only nontrivial graph giving
contribution to $G_{41}$ has no divergent subgraphs, the essential
singular part of the bare propagator is $D_{41}=-c_{41}.$ From the
requirement of vanishing of all poles in
$\langle\Phi\bar{\Phi}\rangle_{B}$ one has
\begin{eqnarray}\label{a}
\widehat P_{44}g^2\left((\overline{a} \alpha_{01}^{(3)} + a
\overline{\alpha}_{01}^{(3)}) + (\overline{a}_2 \alpha_{02}^{(3)}
+ a_2 \overline{\alpha}_{02}^{(3)}) + (\overline{a}_3
\alpha_{03}^{(3)} + a_3 \overline{\alpha}_{03}^{(3)})\right)
-d_2\widehat G_{41}=0.
\end{eqnarray}
After calculating the value of $\widehat P_{44}$  from the pole
equations we find
\begin{equation}\label{al}
d_1^4[(\overline{a} \alpha_{01}^{(3)} + a
\overline{\alpha}_{01}^{(3)}) + (\overline{a}_2 \alpha_{02}^{(3)}
+ a_2 \overline{\alpha}_{02}^{(3)}) + (\overline{a}_3
\alpha_{03}^{(3)} + a_3 \overline{\alpha}_{03}^{(3)})] =
d_2\frac{G_{41}^\Sigma}{9}.
\end{equation}

To reach total finiteness, one can use the remaining coefficients.
From the requirement that $ Z_{2}^{-1}=1$ in four loops one gets, as
in~\cite{KB},
\begin{eqnarray}\label{4lconf}
d_1^3[( \overline{a}_1 \alpha_{21}^{(2)} + a_1
\overline{\alpha}_{21}^{(2)}) + ( \overline{a}_2 \alpha_{22}^{(2)}
+ a_2 \overline{\alpha}_{22}^{(2)}) + ( \overline{a}_3
\alpha_{23}^{(2)} + a_3 \overline{\alpha}_{23}^{(2)})] &=& -
\frac{2d_2}{3} G_{41}^\Sigma,
\\ \nonumber d_1^2[( \overline{a}_1 \alpha_{41}^{(1)} + a_1 \overline{\alpha}_{41}^{(1)}) + (
\overline{a}_2 \alpha_{42}^{(1)} + a_2
\overline{\alpha}_{42}^{(1)}) + ( \overline{a}_3 \alpha_{43}^{(1)}
+ a_3 \overline{\alpha}_{43}^{(1)})] &=& 2d_2G_{41}^\Sigma.
\end{eqnarray}

Again we have the finite and conformal theory up to four loops if
the renormalized Yukawa couplings are chosen to satisfy the
condition
\begin{eqnarray}
&&\sum_{i,k=1}^{3}f_{ik}h_{i}\bar{h}_{k}=|h_{1}|^{2}+|h_2|^2+|h_3|^2=g^{2}
\large{\{} 2+\frac{5}{18}\zeta_5G_{41}^{\Sigma}\varepsilon^{3}
+\frac{5}{3}\zeta_5G_{41}^{\Sigma}(\frac{g^2N}{16\pi^2})\varepsilon^2 \nonumber \\
&+&
5\zeta_5G_{41}^{\Sigma}(\frac{g^2N}{16\pi^2})^2\varepsilon+10\zeta_5G_{41}^{\Sigma}
(\frac{g^2N}{16\pi^2})^3+...\large{\}}, \nonumber\\
\end{eqnarray}
where $G_{41}^{\Sigma}$ was given above (\ref{G}).  For the bare
couplings one has
\begin{equation}\label{res2}
|h_{1}|^{2}_B+|h_2|^2_B+|h_3|^2_B=g^{2}_B\left\{2+\frac{5}{18}\zeta_5G_{41}^{\Sigma}\varepsilon^{3}+...\right\}.
\end{equation}
This again permits, in particular, the value of $|q|\neq 1$, thus
allowing one to obtain a complex deformation of the ${\cal N}=4$
SYM theory with arbitrary complex $\beta$.

\section{Unitarity transformation}

As was first noticed in \cite{Beren}, considering the full
Leigh-Strassler deformation one can find special points in the
parameter space of $\{h_1,h_2,h_3\}$ at which the theory is unitary
equivalent to the $\beta$-deformed ${\cal N}=4$ SYM theory.

Consider a general unitary matrix $U(3)$ ($UU^{+}=1$). It depends on
$9$ parameters. Three of them are the Euler angles and the other six
are the phases. Similarly to the quark mixing, five of six phases
can be eliminated by the redefinition of the chiral superfields.
What is left has the standard Cabbibo-Kobayashi-Maskawa
form\cite{Unitar}
\begin{center}
$ U=\left(
\begin{array}{ccc}
  c_{1} & c_{3} s_{1} & s_{1} s_{3} \\
 - c_{2} s_{1} & c_{1} c_{3} -e^{i y} s_{2} s_{3}
 & e^{i y} c_{3} s_{2}+c_{1} c_{2} s_{3} \\
 s_{1} s_{2} & -c_{1} c_{3} s_{2}-e^{i y} c_{2}
 s_{3} & e^{i y} c_{2} c_{3}-c_{1} s_{2} s_{3} \\
\end{array}
\right) $
\end{center}
where $s_{i}=\sin (x_{i})$ and $c_{i}=\cos (x_{i})$.

We take now the $\beta$-deformed theory and make an arbitrary
unitary transformation of the fields
\begin{equation}\label{transf}
\Phi_i \ = \ U_{ij}\Psi_j.
\end{equation}

After that we demand the new theory to be of the Leigh-Strassler
type. It means the absence of nondiagonal terms like
$Tr(\Psi_{i}\Psi_{j}\Psi_{j})$ $i \neq j$. In the above-defined
parametrization of the unitary matrix this procedure leads to the
\textbf{full} Leigh-Strassler deformation theory provided the
parameters take on the following values:
\begin{equation}\label{unit}
\left\{
\begin{array}{l}
 x_{1}=\pm \arccos(\frac{1}{\sqrt{3}}) + \pi k ,\\
 x_{2}=\frac{\pi}{4} + \frac{\pi l}{2}, \\
 x_{3}=\frac{\pi}{4} + \frac{\pi m}{2},  \\
     y=\frac{\pi}{2} + \pi n. \end{array}
 \right.
\end{equation}
It should be  mentioned that besides the absence of the mixed terms
we would like also to get the coefficients of $Tr(\Psi_{i}^3)$ to be
equal. Indeed, the absence of non-diagonal terms in our case
automatically leads to the equal coefficients of $Tr(\Psi_{i}^3)$ up
to the phases $e^{i \alpha_{i}}$. However, these phases can be
eliminated by the additional phase rotation of the chiral
superfields $\Psi_{i}\rightarrow e^{- i
\frac{\alpha_{i}}{3}}\tilde{\Psi_{i}}$ and one gets the theory of
exactly the Leigh-Strassler form.

As the result, the superpotential which is obtained from the
$\beta$-deformed SYM theory by unitary transformation (\ref{transf})
with parameters fixed by (\ref{unit})  has the form
 \begin{equation}\label{supot}
  W = i  Tr \left( \tilde{h_{1}} \Psi_1\Psi_2\Psi_3
  - \tilde{h_{2}} \Psi_1\Psi_3\Psi_2  \right) + i \frac{\tilde{h_{3}}}{3}
  \sum_{i=1}^3 Tr(\Psi_i^3),
\end{equation}
  where
\begin{equation}\label{equiv}
\begin{array}{lll}
\left\{
\begin{array}{l}
 \tilde{h_{1}}=i(a - b)\\
 \tilde{h_{2}}=i(a +  b)\\
 \tilde{h_{3}}=2i b \\
\end{array}
\right. & \mbox{or} & \left\{
\begin{array}{l}
 \tilde{h_{1}}=e^{\pm \frac{\pi}{3}}(a -b)\\
 \tilde{h_{2}}=e^{\pm \frac{\pi}{3}}(a + b)\\
 \tilde{h_{3}}=-2i b\\
\end{array}
\right.
\end{array}
\end{equation}
Here the factor $e^{\pm \frac{\pi}{3}}$ has the origin from the
different phases of the $Tr(\Psi_{i}^3)$ term for different $i$.
The parameters $a$ and $b$ are linked with the original couplings
$h_1$ and $h_2$ by
\begin{equation}\label{par}
\left\{
\begin{array}{l}
a=\pm \frac{1}{2}(h_1+h_2),\\
b=\pm \frac{1}{i2\sqrt{3}}(h_1-h_2).
\end{array}
\right.
\end{equation}
The signs in expressions for $a$ and $b$ can be chosen
independently.

The chiral propagators calculated in the full Leigh-Strassler
deformed theory (\ref{supot}) with the couplings chosen as
(\ref{equiv},\ref{par}) will be the same as calculated in the
$\beta$-deformed theory. This provides us with nontrivial check of
the calculations made  in the Leigh-Strassler deformed theory.
Namely, taking expressions (\ref{g31},\ref{g41}) and after making a
substitution (\ref{equiv},\ref{par}) one obtains the known results
for the $\beta$-deformed theory~\cite{RSS,Zanon2}.

\section{Exploring the conformal conditions}

Let us consider the calculated expressions for $G_{31}$ in the
non-planar case and $G_{41}$ in the planar case and try to  find
such values of $(h_1,h_2,h_3)$ when these quantities vanish
meaning that the one-loop conformal condition is valid up to three
or four loops. Knowing that in the case of the real beta
deformation in the planar limit the one-loop conformal condition
is exact we are interested in finding new solutions in the full
Leigh-Strassler deformed theory for which the one-loop conformal
condition is also exact.

First of all, similarly to the $\beta$-deformed theory, we have not
found any solution for vanishing of $G_{31}$ in the nonplanar case
which has a simple form and might  be valid in any order of PT.

In the planar case, on the contrary,  we found two families of
simple solutions of the equation  $G_{41} \ = \ 0$.

Solution \# 1:
\begin{equation}
\left\{
\begin{array}{l}
 \tilde{h_{1}}=g e^{i \alpha}(A-B),\\
 \tilde{h_{2}}=g e^{i \alpha}(A+B),\\
 \tilde{h_{3}}= 2g e^{i \alpha} B,
\end{array}
\right.
\end{equation}
where $A,B,\alpha$ are arbitrary \textit{real} numbers.  The
one-loop conformal condition brings us to the following relation
between $A$ and $B$:
$$B^2=\frac{1-A^2}{3}.$$
If this condition is satisfied,  then $G_{41}=0$  for arbitrary
$\alpha$ and $-1\leq A \leq 1$.

However, it is easy to see that solution \# 1 coincides with the
left part of (\ref{equiv}). This means that the obtained theory is
unitary equivalent to the $\beta$-deformed case and is exactly
conformal in the planar limit.

Solution \# 2:
\begin{equation}
\left\{
\begin{array}{l}
 \tilde{h_{1}}=-g e^{i \alpha},\\
 \tilde{h_{2}}=0,\\
 \tilde{h_{3}}= g e^{i \beta}, \\
  \alpha - \beta\neq \frac{2\pi m}{3}
\end{array}
\right. \ \ \ \mbox{or} \ \ \ \ \ \left\{
\begin{array}{l}
 \tilde{h_{1}}=0,\\
 \tilde{h_{2}}=g e^{i \alpha}0,\\
 \tilde{h_{3}}= g e^{i \beta}, \\
  \alpha - \beta\neq \frac{2\pi m}{3}
\end{array}
\right.
\end{equation}
These two cases are equivalent. For $\alpha - \beta = \frac{2\pi
m}{3}$ the obtained theory is unitary equivalent to the real
$\beta$-deformed one, but for arbitrary real values of $\alpha$ and
$\beta$ this is genuine.

Solution \# 3:
\begin{equation}
\left\{
\begin{array}{l}
 \tilde{h_{1}}=g (A-iB),\\
 \tilde{h_{2}}=g (A+iB),\\
 \tilde{h_{3}}= -4igB, \\
\end{array}
\right.
\end{equation}
where $A$ and $B$ are equal to $A=\pm \frac{1}{2}, B=\pm
\frac{1}{2 \sqrt{3}}$. This solution is unitary equivalent to
solution \# 2.

Thus, the only nontrivial solution that exists in the planar limit
and leads to conformal theory (up to 4 loops at least) corresponds
to the superpotential which can be written in the form
\begin{equation}\label{sup}
{\cal W}=i h   \int d^6z (q Tr\Phi_1\Phi_2\Phi_3\\
  - \frac{1}{q} \sum_{i=1}^3\frac{Tr(\Phi_i^3)}{3} ).
\end{equation}
where $|h|^{2}=g^{2}$ and $|q|=1$. The case $q=e^{i \frac{\pi
n}{3}}$  brings us back to the real $\beta$-deformed theory. In the
next section we consider some properties of this theory.

\section{Exact conformal invariance?}

One may wonder if the theory defined by the superpotential
(\ref{sup}) is exactly conformal in the planar limit when $|q|=1$
precisely like the $\beta$-deformed one. To understand whether the
conformal condition is exhausted by one loop, we consider the
corrections to the chiral propagator being interested in the
phase-dependent ones. Due to the unitary equivalence to real
$\beta$-deformed theory for particular values of the phase the
absence of phase-dependent terms would mean the exact conformal
invariance of the theory.

One can observe that the conformal condition in the planar limit is
related to topology of the chiral diagrams~\cite{topol}. The
one-loop conformal condition stays valid in higher orders when the
diagrams contain the "bubbles" on the lines. The next structure that
might emerge is a triangle, but since the propagators are always
chiral-antichiral such a kind of diagrams is forbidden. The next
structure is the "box" present in the "ladder" type diagrams. It
appears for the first time in four-loops and does not contain a
phase factor in the planar limit. To get the phase factor, one
should consider more complicated polygons.

From the superpotential (\ref{sup}) one can notice that only
phase-dependent structures that can emerge are of the form
$$ (|h_3|^{2})^n (|h_1|^{2})^l [(h_3 \bar{h_1})^{3k}+(\bar{h}_3 h_1)^{3k}], k=0,1,.. . $$
Hence, if $h_1=h q,\ h_3=\frac{h}{q}$, $q=e^{i \gamma}$ the only
phase-dependent contribution looks like
$$ const \times \cos (6 k \gamma) .$$

Since we know that when $q=e^{i \frac{\pi n}{3}}$ the theory is
unitary equivalent to the real $\beta$-deformed one, it should be
exactly conformal for $\gamma=\pi n/3$. This corresponds to
$\cos(6k\gamma)=\cos (2 \pi k n)=1$ for arbitrary $k$ and $n$.
Moreover, it is clear that the substitution
$$ \gamma \rightarrow \gamma + \frac{\pi}{3} $$ does not change anything and if a theory
is  exactly conformal for some $\gamma$, it automatically conformal
for
$$\gamma + \frac{\pi n}{3}.$$
This is similar to the beta deformed case  where such an equivalence
was of the form  $$ \beta \rightarrow \beta + \pi n.$$

So the crucial question is whether it is possible to construct a
diagram which is phase-dependent in the planar limit. This
happened to be not a simple task  for the following reasons:

1. All possible phase-dependent "boxes" are suppressed in the
planar limit. Thus, the possible phase-dependent diagram should
contain more complicated structures.

2. The diagram containing a polygon higher than the "box" in which
all phase-dependence is encoded has many external legs. Hence, to
reduce their number to two in order to get the chiral propagator and
keeping only the planar diagrams, one has to make new "boxes" which
again contain no phases. As the result, at least up to ten loops,
one cannot construct a potentially phase-dependent diagram in the
planar limit. We assume, though we have no rigorous proof yet, that
in the planar limit such a phase-dependent structure does not emerge
in any order of PT.

The extra argument for the exact conformal invariance of the
presented theory comes from the the investigation of the
integrability properties of the one-loop dilatation operator in the
full Leigh-Strassler theory made in~\cite{Bund}. The above suggested
solution corresponds to the points in the parameter space where the
theory was found to be integrable in the planar limit. This seems to
be similar to the $\beta$-deformed case where the exact conformal
condition is accompanied with the integrability~\cite{Roiban}.

Thus our conjecture is that the theory defined by the
superpotential (\ref{sup}) with $|q|=1$ is exactly conformal in
the planar limit.

\section{Conclusion}

We have investigated here the conformal conditions for the full
Leigh-Strassler deformation of the ${\cal N}=4$ SYM theory  both in
the planar and  nonplanar cases. The conformal condition was found
up to four loops in the planar limit  and up to three loops in
non-planar case. We would like to emphasize that the obtained theory
is {\it simultaneously} conformal invariant and finite since these
two requirements are {\it identical}. This can be achieved properly
adjusting the Yukawa couplings order by order in PT. In the
framework of dimensional regularization  this requires the double
series over the gauge coupling $g$ and the parameter of dimensional
regularization $\varepsilon$.

Since in the full Leigh-Strassler deformation of the ${\cal N}=4$
SYM theory there is an extra coupling constant, we have more freedom
in our theory. Thus we looked for the solutions where the one-loop
conformal condition is exact and at the same time which are not
obtainable from the real beta deformation of the ${\cal N}=4$ SYM
theory by unitary transformation.  We did not find such solutions in
the nonplanar case but in the planar limit we found one potentially
interesting solution. We made certain that in the planar limit the
one-loop conformal condition in this case is valid up to ten loops
and we present the arguments that it might also be valid in any
order of PT.

If our conjecture is true, then it will be interesting to understand
the nature of this exact conformal condition from the field theory
side as well as from the point of view of the dual gravity
background. While constructing gravity dual background for the
$\beta$-deformed theory the important role was played by the global
$U(1) \times U(1)$ symmetry of the Lagrangian. The theory presented
here has no continuous global symmetries but at some points of the
parameter space it is unitary equivalent to the $\beta$-deformed
theory. This suggests some common features hidden so far. From this
point of view constructing the dual description would be very
interesting.

\section*{Acknowledgements}

Financial support from RFBR grant \# 05-02-17603 and grant of the
Ministry of Education and Science of the Russian Federation \#
5362.2006.2 is kindly acknowledged.


\begin{thebibliography}{99}
\bibitem{Bershadsky} M.Bershadksy, Z.Kakushadze and C.Vafa, \emph{String expansion as large N expansion of gauge theories}, Nucl.Phys.B {\bf 523} (1998) 59 [hep-th/9803076];\\ M.Bershadsky and
A.Johansen, \emph{Large N limit of orbifold field theories},
Nucl.Phys.B {\bf 536} (1998) 141 [hep-th/9803249]
\bibitem{Armoni}A.Armoni, M.Shifman and G.Veneziano, \emph{ Exact results in nonsupersymmetric large N orientifold field theories}, Nucl.Phys.B
{\bf 667} (2003) 170 [hep-th/0302163]; [hep-th/0403071].
\bibitem{Leigh} R.G. Leigh and M.J. Strassler, \emph{Exactly marginal operators and duality in four-dimensional N=1 supersymmetric gauge theory}, Nucl.Phys. {\bf B447} (1995) 95.
\bibitem{Integr1}J.A.Minashan and K.Zarembo, \emph{The Bethe ansatz for N=4 superYang-Mills}, JHEP 0303, 013 (2003),
hep-th/0212208.
\bibitem{Integr2} N.Beisert, C.Kristjansen and M.Staudacher, \emph{ The Dilatation operator of conformal N=4 superYang-Mills theory.},
Nucl.Phys.B {\bf 664}, 131 (2003), hep-th/0303060; \\ N.Beisert and
M.Staudacher, \emph{ The N=4 SYM integrable super spin chain}, Nucl.Phys.B {\bf 670}, 439 (2003), hep-th/0307042; \\
N.Beisert, \emph{ The $su(2|3)$ dynamic spin chain}, Nucl.Phys.B
{\bf 682}, 487 (2004), hep-th/0310252.
\bibitem{Maldacena} J.M. Maldacena, \emph{The Large N limit of superconformal field theories and supergravity}, Adv.Theor.Math.Phys. {\bf 2} (1998) 231; Int.J.Theor.Phys.
{\bf 38} (1999) 1131;\\
S.S. Gubser, I.R. Klebanov and A.M. Polyakov, \emph{Gauge theory correlators from noncritical string theory}, Phys.Lett. {\bf B428} (1998) 105;\\
E. Witten, \emph{Anti-de Sitter space and holography},
Adv.Theor.Math.Phys. {\bf 2} (1998) 253.
\bibitem{Zeytlin}S.A.Frolov, R.Roiban, A.A.Tseytlin, \emph{Gauge-string duality for superconformal deformations of N=4 super Yang-Mills theory}, JHEP {\bf 0507}
(2005) 045, hep-th/0503192; Nucl.Phys.B{\bf 731} (2005) 1,
hep-th/0507021; \\ C-S.Chu and V.V.Khoze, \emph{String theory dual
of the beta-deformed gauge theory}, JHEP {\bf 0607} (2006) 011,
hep-th/0603207.
\bibitem{fin} S. Mandelstam, \emph{Light Cone Superspace and the Ultraviolet Finiteness of the N=4 Model}, Nucl.Phys. {\bf B213} (1983) 149;\\
P.S. Howe, K.S. Stelle and P.K. Townsend, \emph{Miraculous
Ultraviolet Cancellations in Supersymmetry Made Manifest},
Nucl.Phys. {\bf B236} (1984) 125.
\bibitem{LM} O. Lunin and J. Maldacena, \emph{Deforming field theories with U(1) x U(1) global symmetry and their gravity duals}, JHEP {\bf 0505} (2005) 033.
\bibitem{Kulaxizi}M.Kulaxizi, \emph{Marginal Deformations of N=4 SYM and Open vs. Closed String Parameters}, hep-th/0612160;  \emph{On beta-deformations and noncommutativity.}, hep-th/0610310.
\bibitem{AharKol} O. Aharony, B. Kol, S. Yankielowicz,  \emph{On exactly marginal deformations of N=4 SYM and type IIB supergravity on AdS(5) x S**5}, JHEP {\bf 0206} (2002) 039,
hep-th/0205090v2.
\bibitem{BerenCherkis} A. Fayyazuddin, S. Mukhopadhyay, \emph{ Marginal perturbations of N=4 Yang-Mills as deformations of AdS(5) x S**5}, hep-th/0204056v1.
\bibitem{Freedman}D.Z.Freedman and U.Gursoy, \emph{Comments on the beta-deformed N=4 SYM theory}, JHEP {\bf 0511} (2005) 042,
hep-th/0506128.
\bibitem{Zanon1} S. Penati, A. Santambrogio, D. Zanon, \emph{Penati, Two-point correlators in the beta-deformed N=4 SYM at the next-to-leading order}, JHEP {\bf
0510} (2005) 023, [arXiv:hep-th/0506150];\\ A. Mauri, S. Penati, A.
Santambrogio, D. Zanon, \emph{Exact results in planar N=1
superconformal Yang-Mills theory}, JHEP {\bf 0511} (2005) 024,
[arXiv:hep-th/0507282v2].
\bibitem{Zanon2} F. Elmetti, A. Mauri, S. Penati, A. Santambrogio, and D. Zanon, \emph{Conformal invariance of the planar beta-deformed N=4 SYM theory requires beta real}, JHEP {\bf
0701} (2007) 046, [arXiv:hep-th/0606125];\\ F. Elmetti, A. Mauri, S.
Penati, A. Santambrogio, and D. Zanon, \emph{Real versus complex
beta-deformation of the N=4 planar super Yang-Mills theory}, JHEP 10
(2007) 102, arXiv:0705.1483 [hep-th].
\bibitem{RSS} G.C. Rossi, E. Sokatchev and Y.S. Stanev,\emph{New results in the deformed N=4 SYM theory}, Nucl.Phys.
{\bf B729} (2005) 581;  Nucl.Phys. {\bf B754} (2006) 329.
\bibitem{Zanon4}A. Mauri, S. Penati, M.Pirrone, A. Santambrogio, and D.
Zanon, \emph{On the perturbative chiral ring for marginally deformed
N=4 SYM theories}, JHEP {\bf0608} (2006) 072, hep-th/0605145.
\bibitem{KB}D.I.Kazakov and L.V.Bork, \emph{Conformal invariance = finiteness and beta deformed N=4 SYM theory}, JHEP {\bf 0708} (2007) 071,
arXiv:0706.4245.
\bibitem{Ital}F.Elmetti, A.Mauri, M.Pirrone, \emph{Conformal invariance and finiteness theorems for non-planar beta-deformed N=4 SYM theory}, arXiv:0710.4864
[hep-th].
\bibitem{Razamat}S.S.Razamat, \emph{Marginal Deformations of N=4 SYM and of its Supersymmetric Orbifold Descendants}, hep-th/0204043;\\ O.Aharony and
S.S.Razamat,\emph{Exactly marginal deformations of N=4 SYM and of
its supersymmetric orbifold descendants}, JHEP {\bf 0205} (2002)
029, hep-th/0204045.
\bibitem{Madhu}K.Madhu and S.Govindarajan, \emph{A note on perturbative aspects of Leigh-Strassler deformed N=4 SYM theory}, arXiv:0710.5589 [hep-th].
\bibitem{Parkes}A.J.Parkes, \emph{ Three Loop Finiteness Conditions In N=1 Superyang-Mills}, Phys.Lett.B{\bf 156} (1985) 73; A.J.Parkes and P.C.West,\emph{Finiteness in Rigid Supersymmetric Theories}, Nucl.Phys. {\bf B256} (1985) 340.
\bibitem{Jones}I.Jack, D.R.T.Jones and C.G.North, \emph{N=1 supersymmetry and the three loop anomalous dimension for the chiral superfield.}, Nucl.Phys. {\bf B473} (1996)
308.
\bibitem{non-ren} M.T. Grisaru and W. Siegel, \emph{Supergraphity. 2. Manifestly Covariant Rules and Higher Loop Finiteness}, Nucl.Phys. {\bf B201} (1982) 292.
\bibitem{back} M.T. Grisaru, M. Ro\v{c}ek and W. Siegel, \emph{Improved Methods for Supergraphs}, Nucl.Phys. {\bf B159} (1979) 429.
\bibitem{NSVZ} V.A. Novikov, M.A. Shifman, A.I. Vainshtein, and V.I. Zakharov, \emph{Exact Gell-Mann-Low Function of Supersymmetric Yang-Mills Theories from Instanton Calculus}, Nucl.Phys.
{\bf B229} (1983) 381; Phys.Lett. {bf B139} (1984) 389;\\
M.A. Shifman and A.I. Vainshtein, \emph{Solution of the Anomaly
Puzzle in SUSY Gauge Theories and the Wilson Operator Expansion},
Nucl.Phys. {\bf B277} (1986) 456.
\bibitem{finite}
A.V. Ermushev, D.I. Kazakov and O.V. Tarasov, \emph{Construction of
finite N=1 supersymmetric Yang-Mills theories}, Preprint JINR
E2-85-794, Dubna 1985; \emph{Finite N=1 Supersymmetric Grand Unified
Theories},
Nucl.Phys. {\bf B281} (1987) 72;\\
D.R.T. Jones, \emph{Coupling Constant Reparametrization And Finite Field Theories}, Nucl.Phys. {\bf B277} (1986) 153; \\
R. Oehme, \emph{Reduction And Reparametrization Of Quantum Field Theories}, Prog.Theor.Phys.Suppl. {\bf 86} (1986) 215;\\
C. Lucchesi, O. Piguet and K. Sibold, \emph{Necessary And Sufficient
Conditions For All Order Vanishing Beta Functions In Supersymmetric
Yang-Mills Theories}, Helv.Phys.Acta {\bf 61} (1988) 321; Phys.Lett.
{\bf
B201} (1988) 241;\\
X.D. Jiang and X.J. Zhou, \emph{A Criterion For Existence Of Finite
To All Orders N=1 Sym Theories}, Phys.Rev. {\bf D42} (1990) 2109.
\bibitem{scope} L.V. Avdeev and A.A. Vladimirov, \emph{Dimensional Regularization And Supersymmetry}, Nucl.Phys. {\bf B219} (1983) 262.
\bibitem{K1}D.I. Kazakov, \emph{Finite N=1 Susy Field Theories And Dimensional Regularization}, Phys.Lett. {\bf B179} (1986) 352; Mod.Phys.Lett. {\bf A2}
(1987) 663.
\bibitem{SusyMath} A.F.Ferrari http://fma.if.usp.br/~alysson/SusyMath
\bibitem{FeynCalc} http://www.feyncalc.org/
\bibitem{Beren}David Berenstein, Vishnu Jejjala, Robert G. Leigh, \emph{Marginal and relevant deformations of N=4 field theories and noncommutative moduli spaces of vacua}, Nucl.Phys. {\bf B589} (2000) 196-248,
hep-th/0005087v1.
\bibitem{Unitar}P.H.Ginsparg and S.L.Glashow, \emph{Top Quark Mass and Bottom Quark Decay}, Phys.Rev.Lett. {\bf
50} (1983) 1415; \\ L-L.Chau and W-Y.Keung, \emph{Comments on the
Parametrization of the Kobayashi-Maskawa Matrix}, Phys.Rev.Lett.
{\bf 53} (1984) 1802.
\bibitem{topol}V.V.Khoze, \emph{Amplitudes in the beta-deformed conformal Yang-Mills}, JHEP {\bf 0602} (2006) 040,
hep-th/0512194.
\bibitem{Bund} D. Bundzik, T. Mansson, \emph{ The General Leigh-Strassler deformation and integrability}, JHEP {\bf 0601} (2006) 116,
hep-th/0512093v2;\\ T. Mansson, \emph{ The Leigh-Strassler
Deformation and the Quest for Integrability}, JHEP {\bf 06} (2007)
010, hep-th/0703150v2.
\bibitem{Roiban}N.Beiset and R.Roiban, \emph{Beauty and the twist: The Bethe ansatz for twisted N=4 SYM}, JHEP {\bf 0508} (2005) 039,
hep-th/0505187.

\end{thebibliography}
\end{document}